\documentclass[12pt]{article}
\pdfoutput=1
\usepackage{mathtools}
\usepackage{bm}
\usepackage{float}
\usepackage{relsize}
\usepackage{array}
\usepackage{fullpage}
\usepackage{graphicx}
\usepackage{psfrag}
\usepackage{epsf}
\usepackage{amsfonts}
\usepackage{amsmath}
\usepackage{pstricks}
\usepackage{color}
\usepackage{setspace}
\usepackage{easybmat}
\usepackage{slashed}
\usepackage{amsmath}
\usepackage{amssymb}
\usepackage{graphicx}

\usepackage{verbatim}
\usepackage{caption}
\usepackage{subcaption}
\usepackage{multirow}

\usepackage[square,sort,comma,numbers]{natbib}

\usepackage{tikz}

\usepackage{hyperref}

\newcommand{\ignore}[1]{}

\numberwithin{equation}{section}

\allowdisplaybreaks
\def\timesbox{\hbox{$\scriptscriptstyle\times$}}
\def\ant{ {{\lower 1ex  \timesbox} \atop {\raise 1.5ex  \timesbox}}}

\begin{document}

\title{
{\textbf{Study of phases in a holographic QCD model}}
\author{Varun Sethi\footnote{varun.sethi@hotmail.com}\\
\small{{\em Department of Physics and Astrophysics,
University of Delhi,}} \\
\small{{\em Delhi 110007, India}}\\
}}
\date{\vspace{-5ex}}

\maketitle

\abstract{Witten-Sakai-Sugimoto model is used to study two flavour Yang-Mills theory with large number of colours at finite temperature and in presence of chemical potential for baryon number and isospin. Sources for $U(1)_B$ and $U(1)_3$ gauge fields on the flavour 8-branes are D4-branes wrapped on $S^4$ part of the background. Here, gauge symmetry on the flavour branes has been decomposed as $U(2) \equiv U(1)_B \times SU(2)$ and $U(1)_3$ is within $SU(2)$ and generated by the diagonal generator. We show various brane configurations, along with the phases in the boundary theory they correspond to, and explore the possibility of phase transition between various pairs of phases.}

\newpage

\tableofcontents

\newpage
\nocite{*}

\section{Introduction}\label{intro}

Being a strong/weak duality, AdS/CFT correspondence offers an ideal setting for exploring hot and dense QCD. There have been various attempts towards holographic nuclear matter, through bottom-up as well as top-down approaches. Witten \cite{WiHolQCD1} proposed an approach to study non-supersymmetric Yang-Mills theory starting from an instance of AdS/CFT conjecture involving M-theory in $AdS_7 \times S^4$ and a six-dimensional field theory on M5-branes. It has been understood that flavours can be added in the boundary theory through open string sector of D-branes. Sakai-Sugimoto model \cite{SaSu1,SaSu2} extends Witten's construction by adding, in probe approximation, D8-$\overline{\mbox{D8}}$ brane pairs in $N_c$ D4-brane solution of type IIA SUGRA. D4-D8 strings have colour and flavour quantum numbers on the two ends. \ignore{Since these flavour branes intersect with colour branes, there are chiral quarks in the theory. }These 8-branes extend over all coordinates other than $x_4$, which is compactified over a supersymmetry breaking circle. The boundary theory is $SU(N_c)$ Yang-Mills with fundamental quarks. Though $x_4$ direction comes in picture at higher energy scale, the model shows reasonable agreement with lattice results \cite{Rebhanreview}. It also has a geometrical interpretation of chiral symmetry breaking. For a partial list of some other studies involving holographic nuclear matter, see \cite{KarchnKatz}-\cite{similar}.

The motivation for this work partly comes from \cite{BaSw}, which is a top-down approach to holographic superconductor. For some other work on holographic superconductivity, see \cite{GubserTASI10,0908.0011,bottomupscd}.

Reference \cite{BaSw} has in bulk two coincident D8-$\overline{\mbox{D8}}$ pairs arranged as probes in the background generated by the supergravity solution of $N_c$ D4-branes with one dimension compactified on a circle as in \cite{SaSu1}. \ignore{The dynamics of D8 and anti-D8 branes are assumed to be decoupled. (Though anti-D8 branes play no role in the model, their presence is important for force balance and charge balance.)}The 8-8 strings (strings with ends on D8-branes) transform under U(2) gauge group. In \cite{BLL}, a study of holographic QCD was done at finite chemical potential for baryons. The authors have used wrapped 4-branes on $S^4$ (or equivalently, 4-branes dissolved in D8-branes represented as instantons in the 9-dimensional world-volume theory \cite{SaSu1,WiBaryonVertex}) as the source for the electric potential on the 8-branes, which, to leading order, is the chemical potential in boundary theory. The authors of \cite{BaSw} have used two such sources, in Sakai-Sugimoto model with two D8-branes, to introduce baryon and isospin chemical potentials (corresponding to $U(1)_B$ and $U(1)_3$ where $U(2) \sim U(1)_B\times SU(2)$ and $U(1)_3$ is the subgroup of $SU(2)$ corresponding to the diagonal generator) to understand superconductivity in the boundary.\ignore{See Appendix \ref{} for a calculation to justify this.} Since this $t_3$ charge (where $t_3$ is the diagonal generator of $SU(2)$ group) acts differently on the two D8-branes, they separate with an opening angle in the bulk, as shown in Figure \ref{confinedConfig2}. This amounts to breaking of $SU(2)$ symmetry to $U(1)_3$\ignore{ in the boundary theory, which is large $N_c$ $SU(N_c)$ theory with two flavours (in supergravity approximation)}. A comparison of actions shows that the phase with $t_3$ charge turned on is favoured over the phase without it. The tachyonic instability of this intersecting D8-brane configuration is proposed to be the dual of the pairing instability in superconductors.

The work in \cite{BaSw} described above has been done in confined phase with time coordinate Euclidean and compact with periodicity identified by inverse temperature (equation (\ref{ConfB/g})). \ignore{However, beyond a certain higher temperature, there is a transition such that the roles of the imaginary time circle ($ix^0$) and the supersymmetry breaking circle ($x^4$) are interchanged and the energetically favoured background has a horizon (a black brane) \cite{0604161}. This happens at a temperature $T_{HP} = \frac{1}{2\pi R_{S^1}}$ ($R_{S^1}$ being the radius of $x_4$ circle). This transition is analogous to Hawking-Page transition in asymptotic AdS spacetimes \cite{HawPaAdSBHPhTr}, and corresponds to deconfinement transition in the boundary theory.}However, there is another background, asymptotically similar to the previous one, which has the roles of $x_0$ and $x_4$ circles interchanged, in the sense that the blackening function $f(u)$ multiplies $dx_0^2$. This spacetime has a black brane in the bulk and corresponds to, at the boundary, the deconfined phase. In this work, we aim to explore all the phases of large $N_c$ Yang-Mills theory with two flavours at finite baryon number and isospin chemical potential. Further, the configuration of Figure \ref{confinedConfig2} is not the only way to introduce isospin chemical potential. We introduce another brane configuration with baryon number and isospin chemical potentials in the boundary dual theory (see Figure \ref{confinedConfig3}), albeit with UV asymptotics different from that of the configuration in Figure \ref{confinedConfig2}. With tuning parameters as temperature (in the deconfined phase) and chemical potential of the second D8-brane (in both confined and de-confined phases), we compare the actions for various configurations with same UV behaviour to infer which phase is favoured and explore phase transitions between various phases.

The setup of \cite{BaSw} is not symmetric with respect to branes and anti-branes since only the fields on D8-branes (and not on $\overline{\mbox{D8}}$) are taken into consideration. (However, the presence of these anti-branes is important for force balance and charge balance.) Accordingly, the boundary theory in \cite{BaSw} has bound states of only left-handed flavours. In our case, there are both left and right handed flavours in the boundary theory. One subtlety concerns the presence of chiral symmetry broken phase after deconfinement transition. It is assumed that this is indeed the case and a justification is given later in section \ref{Discussion}. An attempt at analysing phases in the boundary dual of Witten-Sakai-Sugimoto model at finite baryon and isospin chemical potential was made in \cite{similar}. However, there were no sources used for the gauge fields $U(1)_B$ and $U(1)_3$. In present work, D4-branes wrapped on $S^4$ part of the background spacetime source $U(1)_B$ and $U(1)_3$ gauge fields as point-like instantons through the Chern-Simons term in the D8-brane action. However, the interactions between these 4-branes are ignored. The non-Abelian gauge field enters the picture by considering each 8-brane to be a stack of $N_f$ of them. For all other purposes, the stack behaves as a single brane as in \cite{BLL,BaSw}.

Comparing the configurations with same UV behaviour, confined phase analysis shows that the nuclear matter phases (Figures \ref{confinedConfig2} and \ref{confinedConfig3}) are uniformly favoured over vacuum. In the deconfined phase, nuclear matter phases E and F (Figure \ref{deconfinedConfig23}) and chiral symmetry restored phase G (corresponding to the parallel brane configuration in Figure \ref{deconfinedConfig015}) are preferred over vacuum. Comparison of phase F with phase G shows various phase transitions with phase G being favoured at higher temperatures.

We start with confined phase analysis in section \ref{confined}. We first discuss the brane configurations resulting from just the baryon number in the theory in section \ref{BaryonConfined}. In section \ref{IsospinConfined}, we discuss isospin chemical potential and free energy of the corresponding brane configurations. Section \ref{confinedResults} shows results for the confined phase. Section \ref{DeconfinedPhaseSection} shows the analysis for the deconfined phase with section \ref{DeconfinedResults} showing the results. Discussion is given in section \ref{Discussion}.

\section{Analysis in the cigar topology background}\label{confined}

This section is devoted to exploring various brane configurations for the setup in the spacetime background of equation (\ref{ConfB/g}) and analysing their thermodynamics. The $x_4-u$ subspace of this spacetime has a cigar topology as shown in Figure \ref{confinedConfig1}. The geometrical picture consists of D8-$\overline{\mbox{D8}}$ pairs embedded in the 10-dimensional D4-brane background given as
$$ ds^2 = R^2\left\{u^\frac{3}{2} 
( \eta _{\mu \nu} dx^{\mu} dx^{\nu} + f(u) dx_4^2) + u^{-\frac{3}{2}}(\frac{du^2}{f(u)} + u^2 d\Omega_4^2)\right\} $$
with
\begin{equation}\label{ConfB/g}
f(u) = 1 - \frac{u_{KK}^3}{u^3}, ~~~~ e^\phi = g_s u^\frac{3}{4}, ~~~~ dc_3 = \frac{2 \pi N_c R^{4}}{ \Omega_4} \epsilon_4, ~~~ R^3 = \pi g_s N_c (\alpha')^{3/2}.
\end{equation}
Here, $\mu=0,1,2,3$ and all coordinates have been made dimensionless by using appropriate powers of $R$, radius of the background spacetime. Temporal coordinate is compactified on a circle, circumference of which is identified as inverse temperature, $\beta$. $\phi$ is the dilaton field and $c_3$ is the dimensionless three-form field on the D4-brane. $x_4$ direction is compactified on $S^1$ and has a period
\begin{equation}\label{x4Period}
    \delta x_4 = \frac{4\pi}{3}\frac{1}{u_{KK}^{1/2}}.
\end{equation}
The scale of compactification is then $\frac{2\pi}{\delta x_4 \cdot R}$ and Yang-Mills coupling constant at this scale is given by
\begin{equation}
    g_{YM}^2=(2\pi)^2 \frac{g_s\alpha'^{1/2}}{\delta x_4 \cdot R}.
\end{equation}
It is below this scale that the dual theory is effectively large $N_c$ QCD. To avoid $\alpha'$ corrections and loop corrections in string theory one needs to impose the limits where radius of the background spacetime is much larger than string length and string coupling is small. (A detailed analysis of supergravity approximation for this case can be found in \cite{D4/D6}.)

To see how this background corresponds to the confined phase, consider a Wilson loop of heavy quark-antiquark separated by a large distance on the boundary. The tendency of the string connecting quark and antiquark is to minimise its tension, which gets a contribution from $g_{00}g_{11}$. This happens at the lowest value of $u$ making the tension constant, which indicates confinement.

\subsection{Finite baryon density}\label{BaryonConfined}

Let us first consider that the theory on the boundary has only baryon number chemical potential as in \cite{BLL}. As explained in section \ref{intro}, for writing action of this configuration, each D8-brane is considered to be a stack of $N_f$ of them. Since baryon chemical potential is equal (to leading order) to the boundary value of the temporal component of $U(1)_B$ gauge field on the D8-brane, we shall consider only the Abelian gauge field (equation (\ref{NAgfdecomp})) in the DBI action. In the following analysis, we write the action for embedding of only D8-branes. Action for anti-branes can be written on similar lines.

Action for the embedding of $N_f$ D8-branes is
\begin{equation}\label{LTS_D8Confined}
    \tilde{S}^{D8}=  C^{D8} \int du
\left\{u^3 f(u)(x_4'(u))^2 + {\frac{1}{f(u)}} \right\}^\frac{1}{2} 
\left\{ u^5 +  (d(u))^2 \right\}^\frac{1}{2}
\end{equation}
with
$$
C^{D8} = \frac{N_f \Omega_4 V_3 \beta R^5}{(2\pi)^8\alpha'^{9/2}g_s}
$$
and $d$ is the canonical conjugate of the zeroth component of the dimensionless $U(1)$ gauge field on D8-branes
\begin{equation}\label{Agfnorm}
a_M = \frac{2\pi\alpha'A_M}{\sqrt{2} R}
\end{equation}
(where $M$ runs over the coordinates of D8-brane),
\begin{equation}
\label{d_confined}
d(u) \equiv  \mbox{} - \frac{1}{C^{D8}} \frac{\partial {\cal L}^{D8}}{\delta a_0^\prime(u)}.
\end{equation}
See Appendix \ref{action} for more details. $V_3$ the spatial volume of the boundary theory spacetime and $\Omega_4$ is the volume of the $S^{4}$ part of the background. 

\begin{figure}[h]
\hspace{4cm}\includegraphics[width=0.6\textwidth]{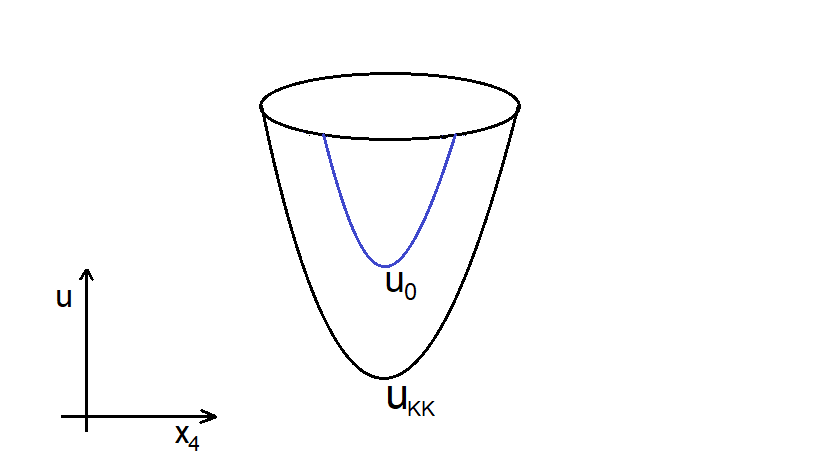}
\caption{Configuration `a' showing the solution of the setup in bulk without sources ($d=0$). Flavour $8-\overline{\mbox{8}}$ branes are shown in blue while the topology of spacetime is in black.}
\label{confinedConfig0}
\end{figure}

\begin{figure}[h]
\hspace{4cm}\includegraphics[width=0.6\textwidth]{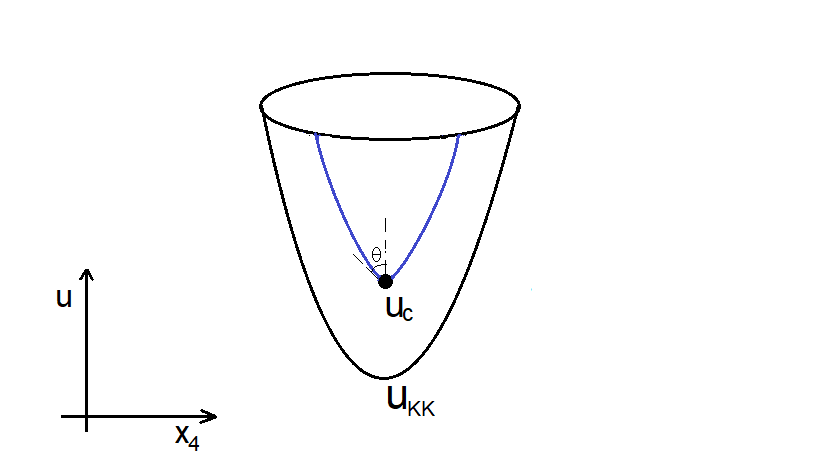}
\caption{Configuration for the bulk setup with sources for $U(1)_B$ field. Flavour $8-\overline{\mbox{8}}$ branes are shown in blue.}
\label{confinedConfig1}
\end{figure}

A non-trivial $d$ can be sourced by instantons of the non-Abelian gauge theory on the 8-brane stack. Part of the action, relevant for us, in the corresponding Chern-Simons term is given by \cite{SaSu1}
\begin{equation}\label{S_CS}
     \frac{N_c}{8\pi^2}\int_{R^4\times R_+} a_0 \mbox{Tr} \hat{F}^2.
\end{equation}
We assume point-like instantons ignoring $\alpha'$ corrections. This assumption is, however, validated by the alternate description of baryons as D4-branes wrapped on $S^4$ part of the background. In this description, action for baryon vertex is the standard DBI action, which evaluates to (see Appendix \ref{action} for formal calculation)
\begin{equation}\label{D4onS4}
   S^{D4} = \frac{1}{3}C^{D8}u_cd.
\end{equation}
Assuming these 4-branes to be smeared over the 3-volume $V_3$ at $u=u_c$,
\begin{equation}\label{NAFlStS_CS}
\mbox{Tr}\hat{F}^2 = 8\pi^2 n_4 \delta(u-u_c) d^3\mathbf{x} du,
\end{equation}
where $n_4$ is the dimensionless density of these 4-branes.
Equations (\ref{LTS_D8Confined}), (\ref{S_CS}) and (\ref{NAFlStS_CS}) give the relevant terms in the action of this theory. Equations of motion are given by
\begin{equation}\label{x4eomConf}
x_4(u):~ (x_4'(u))^2 = \frac{1}{u^3 (f(u))^2}
\left\{\frac{f(u)(u^8+u^3d^2)}{f(u_0)(u_0^8+u_0^3d^2)} - 1 \right\}^{-1}
\end{equation}
and
\begin{equation}\label{eomd}
d(u): d'(u) = \frac{\beta V_3 N_c}{2\pi\alpha' R^2 C^{D8}} n_4 \delta(u-u_c).
\end{equation}
For $d=0$, the brane configuration is shown in Figure \ref{confinedConfig0}.
For a non-zero $d$, it is clear from the geometry of embedding of 8-branes in the background (equation (\ref{ConfB/g})) and the arrangement of 4-branes wrapped on $S^4$ (equation (\ref{NAFlStS_CS})) that the latter appear as a cusp in $x_4-u$ plane as in Figure \ref{confinedConfig1}.
$u_0$ in equation (\ref{x4eomConf}) parametrises the curve; it is the lowest point D8-branes would extend to, if they did not terminate at $u_c$. The asymptotic separation between the D8-$\overline{\mbox{D8}}$ pair can be written as
\begin{equation}\label{asympSeper}
    l = 2\int_{u_c}^\infty du ~x_4'(u).
\end{equation}

As mentioned in section \ref{intro}, the idea is to compare grand canonical potential for various configurations. To study thermodynamics of this system, free energy needs to be calculated. For this purpose, we shall implement baryons as D4-branes wrapped on $S^4$. These 4-branes apply force on the D8-branes, which balances the tension in the latter. The proper tension along D8-brane is given by
\begin{equation}
f^{D8} = C^{D8} u_c^{13/4} \left(1+\frac{d^2}{u_c^5}\right)^{1/2},
\end{equation}
and the force due to D4-branes is given by
\begin{equation}
    f^{D4} = \frac{\partial S^{D4}}{\partial u_c} 
    \left.\frac{1}{\sqrt{g_{uu}}}\right|_{u=u_c}
    = \frac{1}{3} C^{D8} d u_c^{3/4} \sqrt{f(u_c)}.
\end{equation}
For detailed calculation, see Appendix \ref{force}. Equilibrium of this configuration demands
\begin{equation}\label{equilibrium}
f^{D8} \cos\theta = f^{D4},
\end{equation}
where angle $\theta$, in Figure \ref{confinedConfig1}, is given by
\begin{equation}\label{angle}
   \cos\theta = \left\{1 - \frac{g(u_0)}{g(u_c)}\right\}^{1/2},
\end{equation}
with
\begin{equation}\label{gDef}
g(u) \equiv f(u)(u^8+u^3d^2).
\end{equation}
This gives a relation between $u_0$, $u_c$,
\begin{equation}\label{u_0elimConf}
    g(u_0) = g(u_c)\left\{1 - \frac{d^2f(u_c)}{9(d^2+u_c^5)} \right\}.
\end{equation}

\subsection{Introducing isospin chemical potential}\label{IsospinConfined}

We now have two coincident D8-$\overline{\mbox{D8}}$ brane pairs with the wrapped 4-branes appearing as a cusp as in Figure \ref{confinedConfig1}. Let us now discuss how the isospin chemical potential can be implemented in the boundary. We shall review the corresponding brane configuration of \cite{BaSw} and then introduce another one with different UV asymptotics.

\begin{figure}[h]
\hspace{4cm}\includegraphics[width=0.6\textwidth]{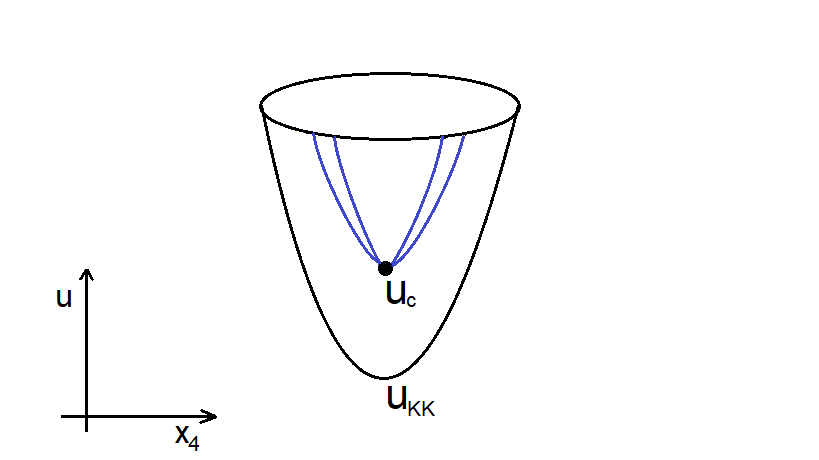}
\caption{Configuration `b' showing a possible arrangement of flavour $8-\overline{\mbox{8}}$ branes (in blue) once the isospin chemical potential is turned on. Flavour branes separate in the UV while crossing each other at the cusp. Topology of spacetime is shown in black.}
\label{confinedConfig2}
\end{figure}

Isospin chemical potential corresponds to the zeroth component of the $U(1)_3$ gauge field on D8-branes at the boundary, the one corresponding to diagonal generator of $SU(2)$. Since this gauge field acts on the two D-branes with opposite sign, the charges on the two D8-branes are $d_1 = d_0 + \frac{d_3}{2}$ and $d_2 = d_0 - \frac{d_3}{2}$ (where $d_0$ is the $U(1)_B$ charge and $d_3$ is the $U(1)_3$ charge). A non-zero value of $d_3$ makes the two D8-branes separate and cross each other in the bulk leading to breaking of the $U(2)$ symmetry to $U(1)_B \times U(1)_3$. (Here, $u_c$ is kept same for both the branes.) The corresponding configuration is shown in Figure \ref{confinedConfig2} \cite{BaSw}.
As stated earlier, for the purpose of instantons, each 8-brane is considered to be a stack of $N_f$ of them. Then, the breaking of symmetry goes as $U(2N_f) \rightarrow U(1) \times U(1)_3 \times SU(N_f) \times SU(N_f) $.

\begin{figure}[h]
\hspace{4cm}\includegraphics[width=0.6\textwidth]{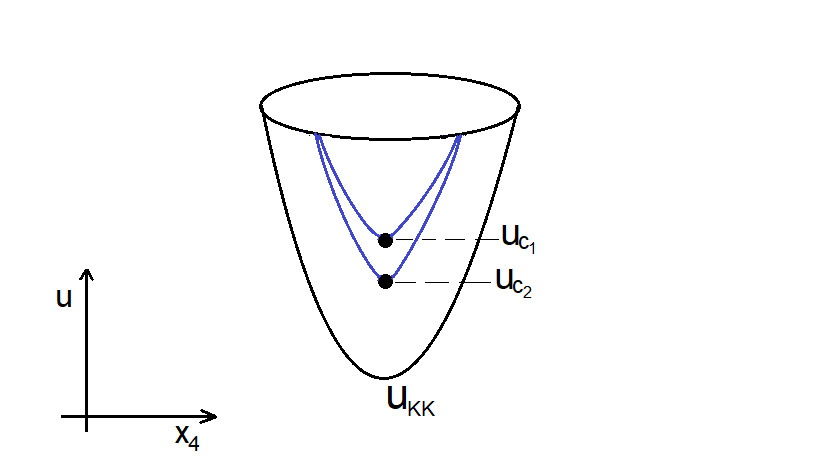}
\caption{Configuration `c' showing a possible arrangement of flavour $8-\overline{\mbox{8}}$ branes (in blue) once the isospin chemical potential is turned on. The branes are asymptotically coincident.}
\label{confinedConfig3}
\end{figure}

Now, we introduce a different brane configuration (shown in Figure \ref{confinedConfig3}) with the boundary theory having baryon and isospin chemical potentials.
It is constructed in such a way that turning on the isospin chemical potential preserves the boundary conditions while the D8-branes separate. This configuration, however, also needs the sources to be separated. Accordingly, we have $u_{c_1}$ different from $u_{c_2}$.

Next, we'd like to study thermodynamics of these separated brane configurations, for which the grand canonical ensemble is a natural choice. Our aim is to evaluate the grand canonical potential function for configurations `b' and `c'. As expected, this is given by the action evaluated at the solution. However, it is convenient to use canonical ensemble for this analysis, the reason for which will be clear later. Accordingly, we have
\begin{equation}\label{grandPot}
    F(n_1,n_2,t) = \Omega(\mu_1,\mu_2,t) + \mu_1 n_1 + \mu_2 n_2.
\end{equation}
Here, $F(n,t)$ is the free energy in the boundary theory, identified with the Legendre transformed action of the bulk configuration
\begin{eqnarray}\label{freeE}
    F(t,d_1,d_2) = \frac{1}{C^{D8}} \left(\tilde{S}_1^{D8}[t,x_{4_1}(u),d_1(u)]_{solution} +  S_1^{D4}(t,d_1,u_{c_1})\right. \nonumber\\
    \left.~~~~ + \tilde{S}_2^{D8}[t,x_{4_2}(u),d_2(u)]_{solution} +  S_2^{D4}(t,d_2,u_{c_2})\right),
\end{eqnarray}
with the subscripts distinguishing the two 8-branes and the corresponding 4-branes of equation (\ref{D4onS4}). $\mu_1$, $\mu_2$ are the baryon chemical potentials corresponding to the two different species of particles with densities proportional to $d_1$ and $d_2$ in the bulk (or $n_1$, $n_2$ in the boundary),
\begin{equation}
    \mu_1 = \left.\frac{\partial F(d_1,d_2,t)}{\partial d_1}\right|_{d_2,t},~~~\mu_2 = \left.\frac{\partial F(d_1,d_2,t)}{\partial d_2}\right|_{d_1,t},
\end{equation}
identified by the boundary value of (a linear combination of) the zeroth component of the $U(1)$ gauge fields on the D8-brane.\ignore{Here, $d_1=d_0+\frac{1}{2}d_3$ and $d_2=d_0-\frac{1}{2}d_3$.}
For the configuration of Figure \ref{confinedConfig2}, we have $u_{c_1}=u_{c_2}$. To calculate $\mu_1$ and $\mu_2$, we make use of the equations written above. Calculation is shown in Appendix \ref{chemicalPotential} and the results are
\begin{equation}
    \mu_{1_b} = \int_{u_c}^{\infty} du \frac{d_1}{\sqrt{f(u) \left(u^5+d_1^2\right) - \frac{g(u_c)}{u^3}  \left\{1 - \frac{d_1^2f(u_c)}{9(d_1^2+u_c^5)}\right\} }}+ \frac{1}{3} u_c,\nonumber
\end{equation}
\begin{equation}\label{chemPotConfinedConf2}
     \mu_{2_b} = \int_{u_c}^{\infty} du \frac{d_2}{\sqrt{f(u) \left(u^5+d_2^2\right) - \frac{g(u_c)}{u^3}  \left\{1 - \frac{d_2^2f(u_c)}{9(d_2^2+u_c^5)}\right\} }}+ \frac{1}{3} u_{c},
\end{equation}
where equation (\ref{u_0elimConf}) has been used to eliminate $u_0$. Using equations (\ref{grandPot}), (\ref{freeE}) and (\ref{chemPotConfinedConf2}), while eliminating $u_0$ using equation (\ref{u_0elimConf}), grand potential for this configuration can be calculated. The result is
\begin{equation}\label{Omega2Conf}
\Omega_b = \int_{u_c}^{\infty} du \left[ \frac{u^{5/2}}{\sqrt{f(u) \left(1 + \frac{d_1^2}{u^5} \right)
- \frac{g(u_c)}{u^8}  \left\{1 - \frac{d_1^2f(u_c)}{9(d_1^2+u_c^5)}\right\}}} + \frac{u^{5/2}}{\sqrt{f(u) \left(1 + \frac{d_2^2}{u^5} \right) 
- \frac{g(u_c)}{u^8}  \left\{1 - \frac{d_2^2f(u_c)}{9(d_2^2+u_c^5)}\right\}}} \right],
\end{equation}

Calculation for chemical potential and grand canonical potential for configuration in Figure \ref{confinedConfig3} goes on the same lines as before, with the modification that $u_{c_1}$ and $u_{c_2}$ are different. The results are
\begin{equation}
    \mu_{1_c} = \int_{u_{c_1}}^{\infty} du \frac{d_1}{\sqrt{f(u) \left(u^5+d_1^2\right) - \frac{g(u_{c_1})}{u^3}  \left\{1 - \frac{d_1^2f(u_{c_1})}{9(d_1^2+u_{c_1}^5)}\right\} }}+ \frac{1}{3} u_{c_1},\nonumber
\end{equation}
\begin{equation}\label{chemPotConfinedConf3}
     \mu_{2_c} = \int_{u_{c_2}}^{\infty} du \frac{d_2}{\sqrt{f(u) \left(u^5+d_2^2\right) - \frac{g(u_{c_2})}{u^3}  \left\{1 - \frac{d_2^2f(u_{c_2})}{9(d_2^2+u_{c_2}^5)}\right\} }}+ \frac{1}{3} u_{c_2}
\end{equation}
and
\begin{equation}\label{Omega3Conf}
\Omega_c = \int_{u_{c_1}}^{\infty} du \frac{u^{5/2}}{\sqrt{f(u) \left(1 + \frac{d_1^2}{u^5} \right)
- \frac{g(u_{c_1})}{u^8}  \left\{1 - \frac{d_1^2f(u_{c_1})}{9(d_1^2+u_{c_1}^5)}\right\}}} +  \int_{u_{c_2}}^{\infty} du \frac{u^{5/2}}{\sqrt{f(u) \left(1 + \frac{d_2^2}{u^5} \right) 
- \frac{g(u_{c_2})}{u^8}  \left\{1 - \frac{d_2^2f(u_{c_2})}{9(d_2^2+u_{c_2}^5)}\right\}}}.
\end{equation}
The corresponding result for configuration `a' in Figure \ref{confinedConfig0} is given by
\begin{equation}\label{Omega0Conf}
    \Omega_a = 2\int_{u_{0}}^{\infty} du \frac{u^{5/2}}{\sqrt{f(u)-\frac{u_0^8}{u^8}f(u_0)}}.
\end{equation}

\subsection{Results}\label{confinedResults}

In the spacetime background of equation (\ref{ConfB/g}), we have seen three configurations. These correspond to various phases in the dual theory on the boundary as shown in Table \ref{confinedConfigPhase}.
\begin{table}[h!]
\begin{center}
\begin{tabular}{ |c|c|c| }
 \hline
 Configuration & Phase identifier & Description of phase \\
 \hline\hline
 \begin{tabular}{@{}c@{}}
    Configuration `a'\\(Figure \ref{confinedConfig0})
 \end{tabular}
 & Phase A & Vacuum \\
 \hline
 \begin{tabular}{@{}c@{}}
    Configuration `b'\\(Figure \ref{confinedConfig2})
 \end{tabular}
 & Phase B & 
   \begin{tabular}{@{}c@{}}
        $U(2)$ symmetry broken to $U(1)_B \times U(1)_3$
    \end{tabular} \\
 \hline
 \begin{tabular}{@{}c@{}}
    Configuration `c'\\(Figure \ref{confinedConfig3})
 \end{tabular}
 & Phase C &
    \begin{tabular}{@{}c@{}}
        $U(2)$ symmetry seemingly restored asymptotically
    \end{tabular} \\
 \hline
\end{tabular}
\end{center}
\caption{Various configurations in bulk and the corresponding phases in the boundary theory in confined phase}
\label{confinedConfigPhase}
\end{table}
The idea is to compare the phases represented by brane configurations with same boundary conditions. All calculations are done for $u_{KK}^3=1/2$ and $l=1$.

\begin{figure}[h!]
\hspace{1.5cm}\includegraphics[width=0.4\textwidth]{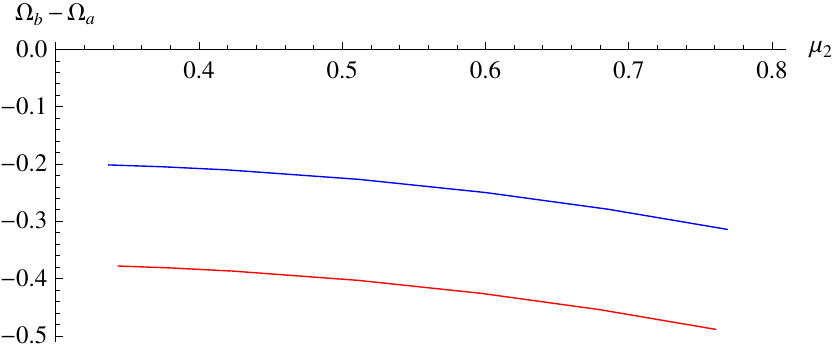}
\hspace{.5cm}\includegraphics[width=0.35\textwidth]{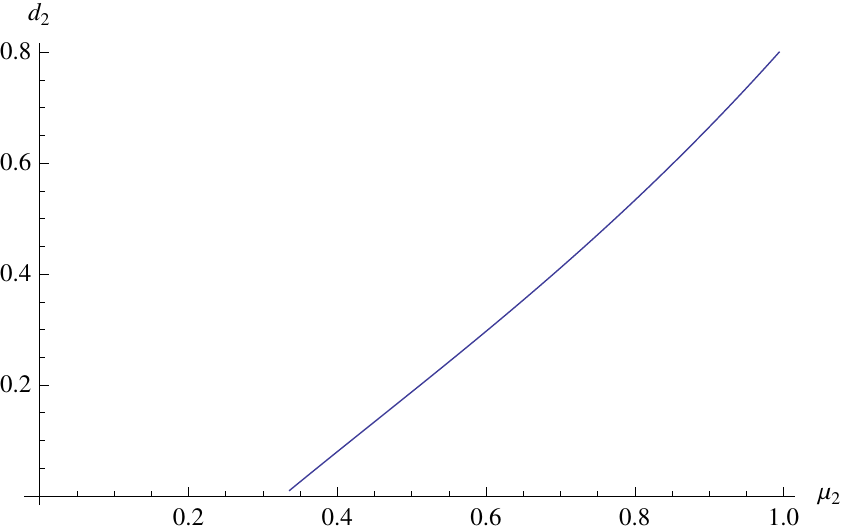}
\caption{$\Omega_{ba}$ \mbox{vs} $\mu_2$. Red and blue curves are plotted at $d_{1b}=1$ and $d_{1b}=.7$ respectively. Phase B is preferred over phase A (vacuum).}
\label{O20confined}
\end{figure}

Let first consider phases A and B . Figure \ref{O20confined} shows the plots of $\Omega_{ba}(=\Omega_b-\Omega_a)$ vs $\mu_2$ with $d_{1b}$ held fixed at $1$ (in red), at $.7$ (in blue) and of $\mu_2$ vs $d_{2b}$. (The first subscript in $d_{1b}$ stands for first 8-brane and the second one is for configuration b.) $\mu_2$ is varied by varying $d_{2b}$, which is taken from $0$ to $d_{1b}$ so that the difference ($d_{3b}=d_{b1}-d_{2b}$), remains positive. We observe that phase B is uniformly preferred over vacuum.

\begin{figure}[h!]
\centering
\includegraphics[width=0.4\textwidth]{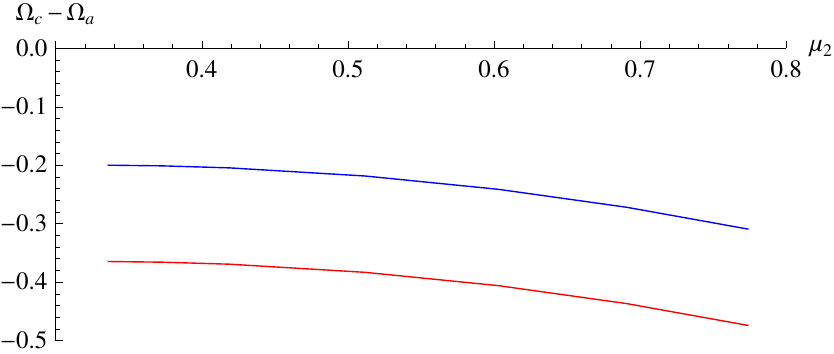}
\caption{$\Omega_{ca}$ \mbox{vs} $\mu_2$. Red and blue curves are plotted at $d_{1c}=1$ and $d_{1c}=.7$ respectively. Phase C is preferred over phase A (vacuum).}
\label{O30confined}
\end{figure}

We look at the comparison of phases A and C \ignore{(Figures \ref{confinedConfig0} and \ref{confinedConfig3})}now. Figure \ref{O30confined} shows the plots of $\Omega_{ca}$ vs $\mu_2$ for $d_{1c}=1$ (red) and $d_{1c}=.7$ (blue). We observe that phase C is also uniformly preferred over vacuum.



All these results are summarised in the table \ref{confinedResultsSummary}.
\begin{table}[h!]
\centering
\begin{tabular}{|c||c|c|c|}
    \hline
      & $\Omega_a$ & $\Omega_b$ & $\Omega_c$\\
    \hline\hline
    $\Omega_a$ & $=0$ & $>0$ & $>0$\\
    \hline
    $\Omega_b$ & $<0$ & $=0$ & -\\
    \hline
    $\Omega_c$ & $<0$ & - & $=0$\\
    \hline
\end{tabular}
\caption{Summary of confined phase results\ignore{ shown in Figures \ref{O20confined}, \ref{O30confined} and \ref{O23confined}}. Results in various boxes show quantity in first row subtracted out of the quantity in first coloumn.}
\label{confinedResultsSummary}
\end{table}

\section{Analysis in black brane background}\label{DeconfinedPhaseSection}

Let us now explore various configurations for this system in the spacetime of equation (\ref{DeconfB/g}). Background metric in this phase is given as
$$ ds^2 = R^2\left\{u^\frac{3}{2} 
(f(u)dx_0^2 + dx_{i}^2 + dx_4^2) + u^{-\frac{3}{2}}(\frac{du^2}{f(u)} + u^2 d\Omega_4^2)\right\}, $$
with
\begin{equation}\label{DeconfB/g}
    f(u) = 1 - \frac{u^3_T}{u^3},~~
    u_T = \left(\frac{4\pi}{3}\right)^2 t^2, ~~ t = RT.
\end{equation}
$t$ is the dimensionless temperature. Since there is a horizon, the quark-antiquark string tension (mentioned in the confined phase) vanishes, signalling deconfinement.

\begin{figure}[h]
\hspace{3cm}\includegraphics[width=0.6\textwidth]{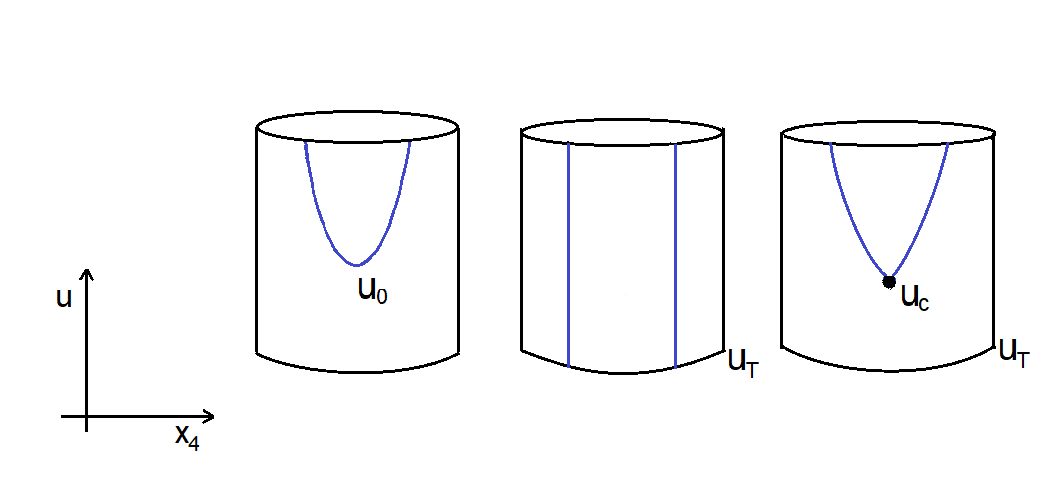}
\caption{\ignore{From left to right, configuration 'd', configuration 'g' and configuration 1;} Brane configurations for the bulk setup in deconfined phase with only baryon chemical potential. Flavour $8-\overline{\mbox{8}}$ branes are shown in blue while the topology of spacetime is in black. Presence of sources does not change parallel branes configuration.}
\label{deconfinedConfig015}
\end{figure}

The action for D8-brane embedding is given by
\begin{equation}\label{S_D8Deconfined}
    S^{D8} = {C^{D8}}
    \int du~ 
    \left\{u^8 f(u) (x_4^\prime(u))^2 
    + u^5 
    \left( 1 - (a_0^\prime(u))^2 \right)
\right\}^\frac{1}{2},
\end{equation}
which after Legendre transform becomes
\begin{equation}\label{LTS_D8Deconfined}
    \tilde{S}^{D8}=  C^{D8} \int du
    \left\{u^3 f(u)(x_4'(u))^2 + 1 \right\}^\frac{1}{2} 
    \left\{u^5 + (d(u))^2 \right\}^\frac{1}{2}.
\end{equation}
Here, $d$ is given by
\begin{equation}
    d(u) = \frac{u a_0'(u)}{\left[ f(u) (x_4'(u))^2 
    + u^{-3} \left(1 - ({a}_0'(u))^2 \right) \right]^{\frac{1}{2}}}.
\end{equation}
First, we consider only baryon number density in the boundary theory. Just as in the confined phase, the source for non-trivial $d$ can be wrapped 4-branes as instantons on the D8-branes. \footnote{In the deconfined phase, strings stretching between the 8-branes and the horizon can also source the gauge field. However, it has been shown in \cite{BLL} that the corresponding brane configuration is unstable to density fluctuations.} Accordingly, the equation of motion for $d$ remains the same as in equation (\ref{eomd}). For $x_4(u)$, there are two possible solutions,
$$ x_4'(u)=0 $$
and
$$ (x'_4(u))^2 = \frac{1}{u^3 f(u)}
\left\{\frac{f(u)(u^8+u^3d^2)}{f(u_0)(u_0^8+u_0^3d^2)} - 1 \right\}^{-1}. $$
Various brane configurations, with and without sources, are shown in Figure \ref{deconfinedConfig015}.
The dimensionless asymptotic separation between the D8-$\overline{\mbox{D8}}$ pair is again
\begin{equation}\label{asympSeper}
    l = 2\int_{u_c}^\infty du ~x_4'(u).
\end{equation}
However, in this case, there are two values of $u_c$ for each $l$, upto $l_{max}$, beyond which there is no solution. Accordingly, there are two "cusp" (or "U", for $d=0$) solutions.

The analysis for force balance goes on the same lines as in the confined phase. The proper tension in D8-brane can be calculated using the derivative of the action equation (\ref{LTS_D8Deconfined}) wrt $u_c$, on the lines of appendix \ref{force}. This is given as
\begin{equation}
f_{D8} = C^{D8} u_c^{13/4} \sqrt{f(u_c)} \left(1 + \frac{d^2}{u_c^5}\right)^{1/2}.
\end{equation}
Action for the D4-branes wrapped on $S^4$ part of the background evaluates to
\begin{equation}
S_{D4} = \frac{1}{3} C^{D8} u_c \sqrt{f(u_c)}~d,
\end{equation}
which gives
\begin{equation}
    f_{D4} = \frac{1}{3} C^{D8} d~ \frac{3-f(u_c)}{2} u_c^{3/4}
\end{equation}
for the force due to D4-branes. The relation between $u_0$ and $u_c$ now reads
\begin{equation}\label{u_0elimDeconf}
    g(u_0) = g(u_c)\left\{1 - \frac{d^2(3-f(u_c))^2}{36f(u_c)(d^2+u_c^5)} \right\}.
\end{equation}

\begin{figure}[h]
\hspace{4cm}\includegraphics[width=0.6\textwidth]{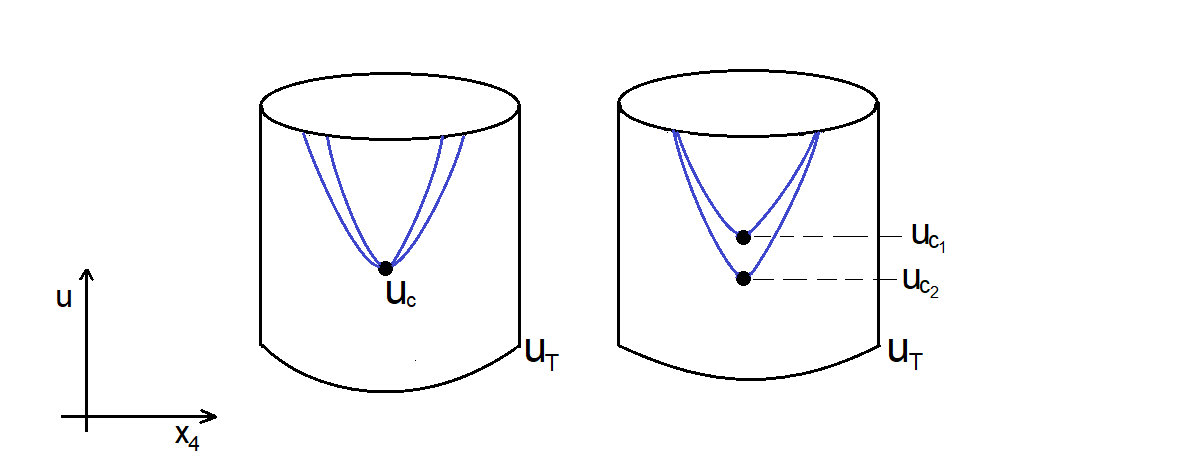}
\caption{Configurations `e' and `f', showing possible arrangement of flavour $8-\overline{\mbox{8}}$ branes (in blue) in the deconfined phase of the theory with baryon number and isospin chemical potential.}
\label{deconfinedConfig23}
\end{figure}

Once the isospin chemical potential is turned on, the two D8-branes will separate, as explained in section \ref{IsospinConfined}. Again, we have two possible curved brane configurations with different UV asymptotic behaviour as shown in Figure \ref{deconfinedConfig23}.
For parallel brane configuration, giving different densities to different branes does not make them separate any further than they already are, since their profile is independent of density. The analysis for calculating chemical potential for the two 8-branes for the confined phase in section \ref{confined} also holds for deconfined phase. The results are
\begin{equation}
    \mu_{1_e} = \int_{u_c}^{\infty} du \frac{d_1\sqrt{f(u)}}{\sqrt{f(u) \left(u^5+d_1^2\right) - \frac{g(u_c)}{u^8} \left\{1 - \frac{d_1^2(3-f(u_c))^2}{36f(u_c)(d_1^2+u_c^5)} \right\} }}+ \frac{1}{3} u_c\sqrt{f(u_c)},\nonumber
\end{equation}
\begin{equation}\label{chemPotDeconfinedConf2}
     \mu_{2_e} = \int_{u_c}^{\infty} du \frac{d_2\sqrt{f(u)}}{\sqrt{f(u) \left(u^5+d_2^2\right) - \frac{g(u_c)}{u^8} \left\{1 - \frac{d_2^2(3-f(u_c))^2}{36f(u_c)(d_2^2+u_c^5)} \right\} }}+ \frac{1}{3} u_{c}\sqrt{f(u_c)}
\end{equation}
for configuration `e' and
\begin{equation}
    \mu_{1_f} = \int_{u_{c_1}}^{\infty} du \frac{d_1\sqrt{f(u)}}{\sqrt{f(u) \left(u^5+d_1^2\right) - \frac{g(u_{c_1})}{u^8} \left\{1 - \frac{d_1^2(3-f(u_{c_1}))^2}{36f(u_{c_1})(d_1^2+u_{c_1}^5)} \right\} }}+ \frac{1}{3} u_{c_1}\sqrt{f(u_{c_1})},\nonumber
\end{equation}
\begin{equation}\label{chemPotDeconfinedConf3}
     \mu_{2_f} = \int_{u_{c_2}}^{\infty} du \frac{d_2\sqrt{f(u)}}{\sqrt{f(u) \left(u^5+d_2^2\right) - \frac{g(u_{c_2})}{u^8} \left\{1 - \frac{d_2^2(3-f(u_{c_2}))^2}{36f(u_{c_2})(d_2^2+u_{c_2}^5)} \right\} }}+ \frac{1}{3} u_{c_2}\sqrt{f(u_{c_2}})
\end{equation}
for configuration `f' in Figure \ref{deconfinedConfig23}, and
\begin{equation}
    \mu_{1_g} = \int_{u_T}^{\infty} du \frac{d_1}{\sqrt{u^5+d_1^2}},\nonumber
\end{equation}
\begin{equation}\label{chemPotDeconfinedConf5}
    \mu_{2_g} = \int_{u_T}^{\infty} du \frac{d_2}{\sqrt{u^5+d_2^2}}
\end{equation}
for parallel branes configuration (`g') in Figure \ref{deconfinedConfig015}. Note, that this parallel brane configuration can have non-zero $d$ without sources since the strings from D8-branes can end at the horizon. Grand canonical potentials for these solutions can be calculated using equations (\ref{grandPot}), (\ref{freeE}), (\ref{chemPotDeconfinedConf2}), (\ref{chemPotDeconfinedConf3}), (\ref{chemPotDeconfinedConf5}) and (\ref{u_0elimDeconf}). The expressions are written next.
\begin{eqnarray}\label{Omega2Deconf}
\Omega_e = \int_{u_c}^{\infty} du \left[ \frac{u^{5/2}\sqrt{f(u)}}{\sqrt{f(u) \left(1 + \frac{d_1^2}{u^5} \right)
- \frac{g(u_c)}{u^8} \left\{1 - \frac{d_1^2(3-f(u_c))^2}{36f(u_c)(d_1^2+u_c^5)} \right\}}} \right. \nonumber \\
\left.~~~ + \frac{u^{5/2}\sqrt{f(u)}}{\sqrt{f(u) \left(1 + \frac{d_2^2}{u^5} \right) 
- \frac{g(u_c)}{u^8}  \left\{1 - \frac{d_2^2(3-f(u_c))^2}{36f(u_c)(d_2^2+u_c^5)} \right\}}} \right].
\end{eqnarray}
\begin{eqnarray}\label{Omega3Deconf}
\Omega_f = \int_{u_{c_1}}^{\infty} du \frac{u^{5/2}\sqrt{f(u)}}{\sqrt{f(u) \left(1 + \frac{d_1^2}{u^5} \right)
- \frac{g(u_{c_1})}{u^8}  \left\{1 - \frac{d_1^2(3-f(u_{c_1}))^2}{36f(u_{c_1})(d_1^2+u_{c_1}^5)} \right\}}} + \\ ~~~~~\int_{u_{c_2}}^{\infty} du \frac{u^{5/2}\sqrt{f(u)}}{\sqrt{f(u) \left(1 + \frac{d_2^2}{u^5} \right) 
- \frac{g(u_{c_2})}{u^8}  \left\{1 - \frac{d_2^2(3-f(u_{c_2}))^2}{36f(u_{c_2})(d_2^2+u_{c_2}^5)} \right\}}}.
\end{eqnarray}
\begin{equation}\label{Omega5Deconf}
    \Omega_g = \int_{u_T}^{\infty} du \left( \frac{u^5}{\sqrt{u^5+d_1^2}} + \frac{u^5}{\sqrt{u^5+d_2^2}} \right).
\end{equation}
The corresponding expression for the U-shaped configuration (`d') in Figure \ref{deconfinedConfig015} is
\begin{equation}\label{Omega0Deconf}
    \Omega_d = 2\int_{u_{0}}^{\infty} du \frac{u^{5/2}\sqrt{f(u)}}{\sqrt{f(u)-\frac{u_0^8}{u^8}f(u_0)}}.
\end{equation}

\subsection{Results}\label{DeconfinedResults}

Various phases corresponding to the configurations seen in this section are summarised in Table \ref{deconfinedConfigPhase}.
\begin{table}[h!]
\begin{center}
\begin{tabular}{ |c|c|c| }
 \hline
 Configuration & Phase identifier & Description of phase \\
 \hline\hline
 \begin{tabular}{@{}c@{}}
    Configuration `d' (U-shaped)\\(Figure \ref{deconfinedConfig015})
 \end{tabular}
 & Phase D & Vacuum \\
 \hline
 \begin{tabular}{@{}c@{}}
    Configuration `e'\\(1st diagram in Figure \ref{deconfinedConfig23})
 \end{tabular}
 & Phase E & 
   \begin{tabular}{@{}c@{}}
        $U(2)$ symmetry broken to $U(1)_B \times U(1)_3$
    \end{tabular} \\
 \hline
 \begin{tabular}{@{}c@{}}
    Configuration `f'\\(Figure \ref{deconfinedConfig23})
 \end{tabular}
 & Phase F &
    \begin{tabular}{@{}c@{}}
        $U(2)$ symmetry seemingly restored asymptotically
  \end{tabular} \\
 \hline
 \begin{tabular}{@{}c@{}}
    Configuration `g' (parallel brane)\\(Figure \ref{deconfinedConfig015})
 \end{tabular}
 & Phase G &
    \begin{tabular}{@{}c@{}}
        U(2) symmetry restored
  \end{tabular} \\
  \hline
\end{tabular}
\end{center}
\caption{Various configurations in bulk and the corresponding phases in the boundary theory in confined phase}
\label{deconfinedConfigPhase}
\end{table}
Our task is to thermodynamically compare these phases. We can already infer, from equation (\ref{Omega5Deconf}), that for parallel brane configuration, non-zero $d$ is preferred over $d=0$. Hence, parallel brane configuration with $d=0$ can not represent vacuum phase of the boundary theory and it need not be considered. For numerics, we again keep $l=1$ and consider the solution with larger value of $u_c$ (or $u_0$) as in \cite{BLL} since it lies lower on the energy scale than the solution with smaller $u_c$.

Let us first compare phases D and E.
\begin{figure}[h!]
\hspace{1.2cm}\includegraphics[width=0.4\textwidth]{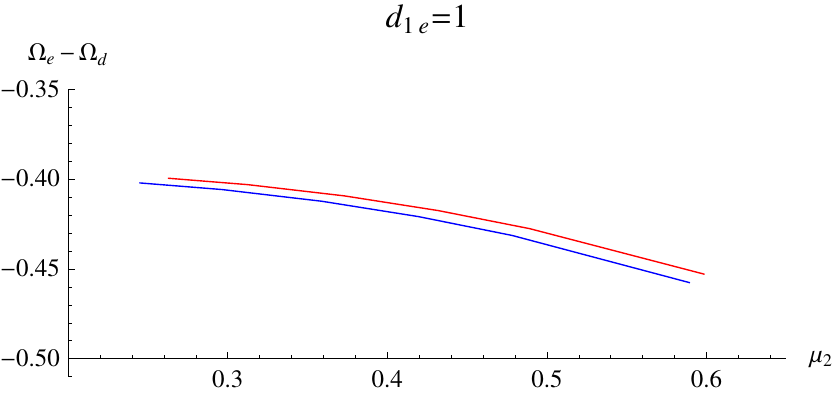}
\hspace{.3cm}\includegraphics[width=0.4\textwidth]{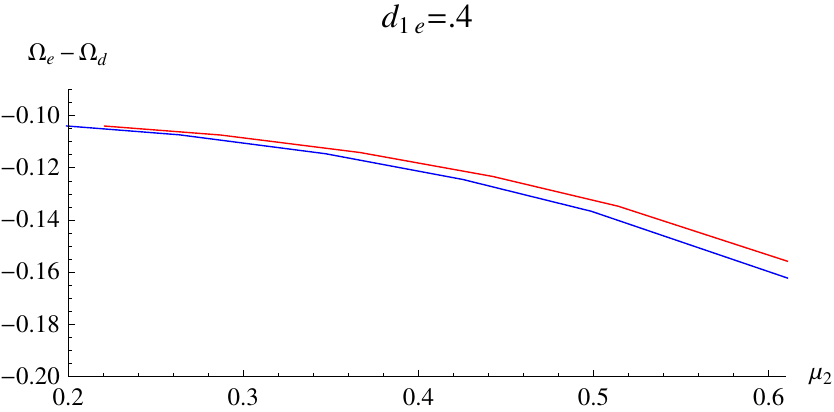}\\
\caption{$\Omega_{ed}$ \mbox{vs} $\mu_2$. Red and blue plots are at $t=.1$ and $t=.14$ respectively.}
\label{O20Deconfined}
\end{figure}
\begin{figure}[h!]
\centering
\includegraphics[width=0.4\textwidth]{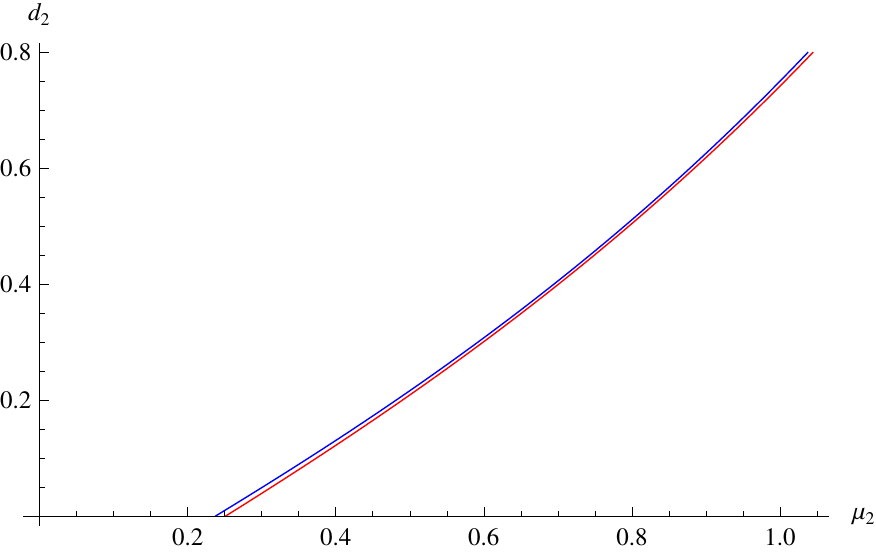}
\caption{$d_2$ vs $\mu_2$ at $t=.1$ (red) and $t=.135$ (blue).}
\label{d2mu2Deconfinedtispt1}
\end{figure}
Figures \ref{O20Deconfined} and \ref{d2mu2Deconfinedtispt1} show the plots for $\Omega_{ed}$ vs $\mu_2$ and $\mu_2$ vs $d_2$, with $d_{1e}$ held fixed at $1$ in the first plot and at $.4$ in the second plot of Figure \ref{O20Deconfined}. It should be noted that though red and blue curves are at the same value of $d_{1e}$, $\mu_1$ for the two plots is not same. This happens because on changing temperature, the configuration itself changes. Accordingly, $\Omega$ changes not only with temperature directly, but also indirectly. Phase E is uniformly preferred over phase D.

\begin{figure}[h!]
\centering
\includegraphics[width=0.4\textwidth]{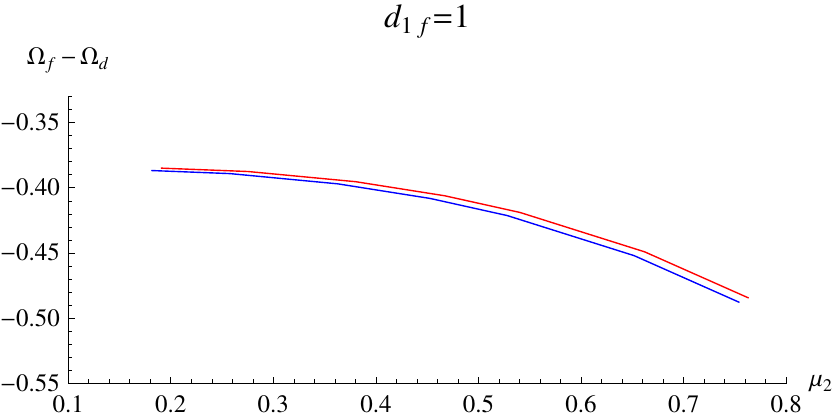}
\hspace{.3cm}\includegraphics[width=0.4\textwidth]{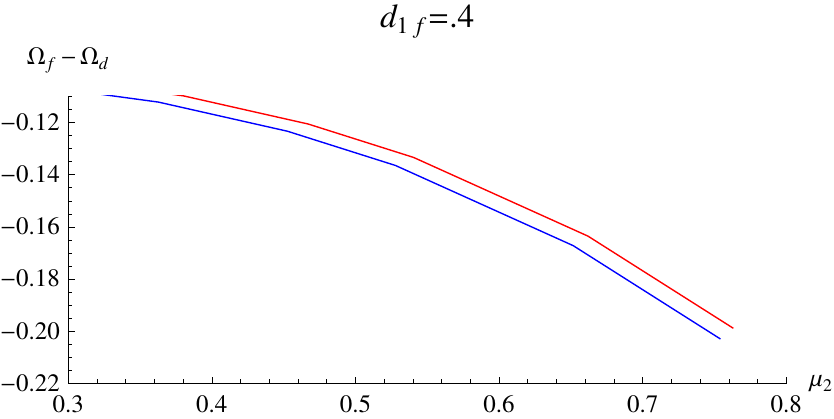}\\
\caption{$\Omega_{fd}$ \mbox{vs} $\mu_2$. Red and blue plots are at $t=.1$ and $t=.13$ respectively.}
\label{O30Deconfined}
\end{figure}

Figure \ref{O30Deconfined} shows plots for the comparison of phases D and F with $d_{1f}=1$ and $d_{1f}=.4$ respectively.
Phase F is preferred over phase D with no phase transition.

\begin{figure}[h!]
\centering
\includegraphics[width=0.4\textwidth]{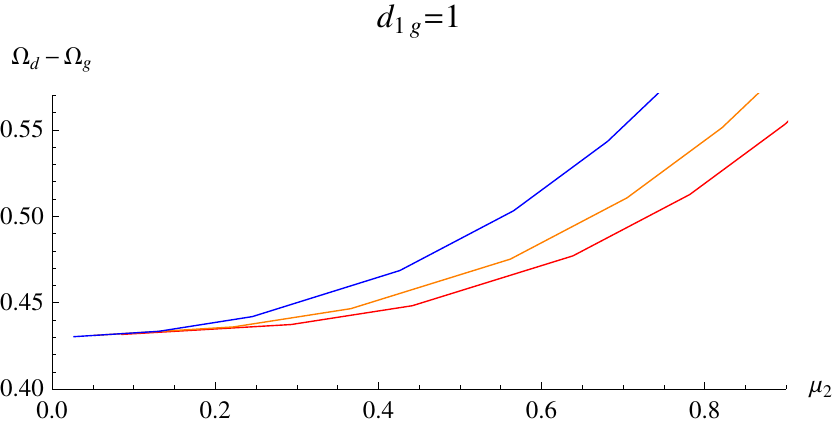}
\includegraphics[width=0.4\textwidth]{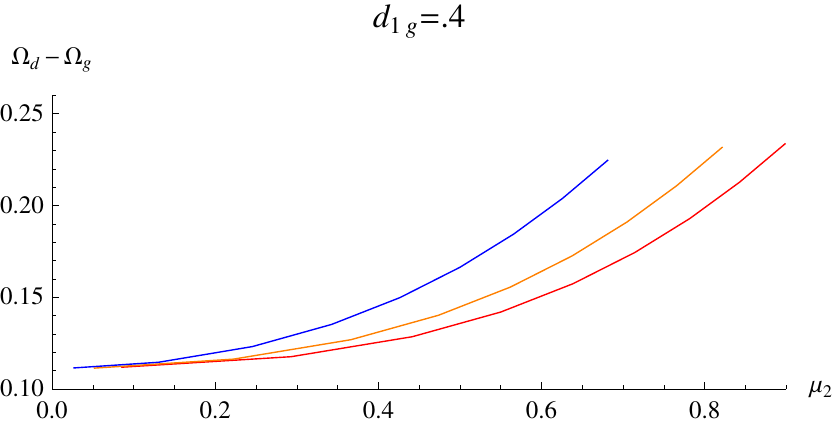}\\
\caption{$\Omega_{dg}$ \mbox{vs} $\mu_2$. Red, orange and blue plots are at $t=.1$, $t=.12$ and $t=.15$ respectively.}
\label{O05Deconfined}
\end{figure}

Next, we compare phases D and G. Plots are shown in Figure \ref{O05Deconfined}.
There is no phase transition and phase G is preferred over phase D.

\begin{figure}[h!]
\centering
\includegraphics[width=0.4\textwidth]{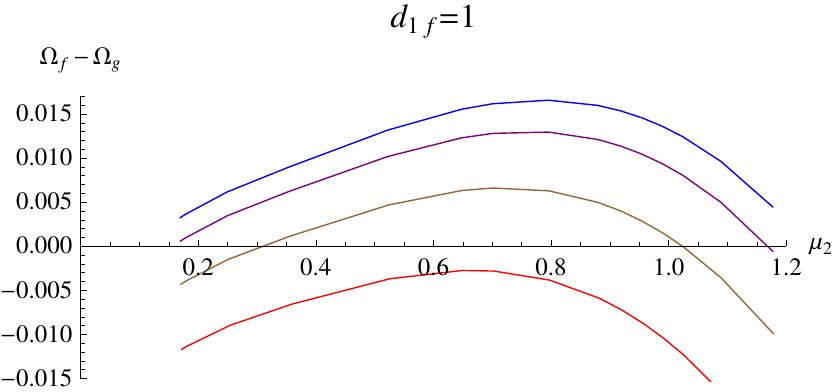}
\caption{$\Omega_{fg}$ \mbox{vs} $\mu_2$ for $d_1=1$. Red, brown, purple and blue curves are at $t=.133$, $t=.135$, $t=.1363$ and $t=.137$ respectively.}
\label{O35Deconfd1is1}
\end{figure}

\begin{figure}[h!]
\centering
\includegraphics[width=0.4\textwidth]{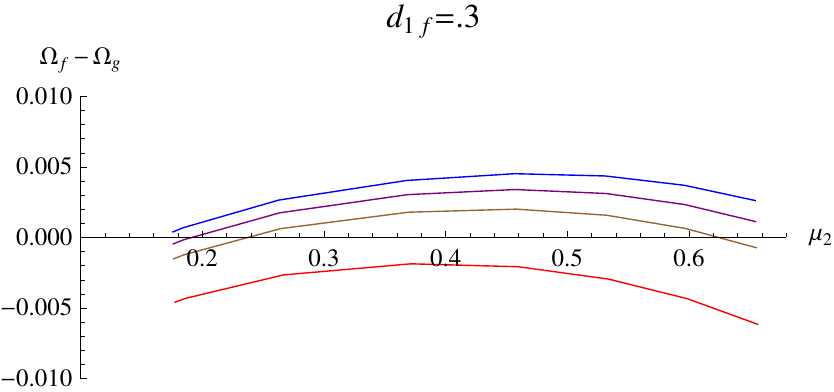}\\
\caption{$\Omega_{fg}$ \mbox{vs} $\mu_2$ for $d_1=.3$. Red, brown, purple and blue curves are at $t=.12$, $t=.123$, $t=.124$ and $t=.1248$ respectively. Note, that the pattern has turned anti-clockwise compared to that in Figure \ref{O35Deconfd1is1}.}
\label{O35Deconfd1ispt3}
\end{figure}

\begin{figure}[h!]
\centering
\includegraphics[width=0.4\textwidth]{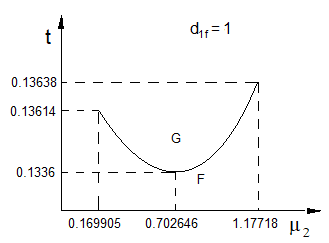}
\includegraphics[width=0.4\textwidth]{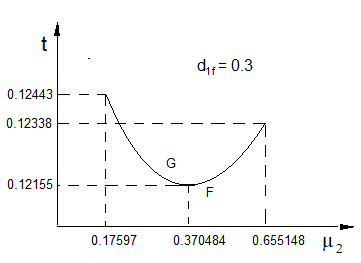}
\caption{Phase diagram showing phases F and G.}
\label{phaseFG}
\end{figure}

Figures \ref{O35Deconfd1is1} and \ref{O35Deconfd1ispt3} show plots of $\Delta\Omega_{fg}$ (phases F and G) vs $\mu_2$ for various $t$ and $d_{1f}$ values. As before, to vary $\mu_2$, $d_{2f}$ is varied from $0$ to $d_{1f}$. Shape of the plots and behaviour with temperature is in line with the plots for nuclear vs QGP phases in \cite{BLL}. This is expected since configurations `f' and `g' are analogous to cusp and parallel brane configurations of \cite{BLL}. At low temperatures, phase F is favoured over the phase G. Then there is a range of higher temperature with re-entrant phase transition, that is phase F to phase G and then back to phase F. As temperature is raised further, we see that the curve starts at a positive value, implying that phase G is favoured over phase F. However, for a large value $\mu_2$, there is a transition to phase F. For temperatures higher than this, phase G is favoured over phase F. On reducing the value of $d_{1f}$, we see that the whole pattern shifts anti-clockwise and the curves are closely spaced implying that the ranges of temperatures for which there are phase transitions become smaller. Caricatures of the corresponding phase diagrams with the coordinates of the end points are shown in Figure \ref{phaseFG}.

These results are summarised in the Table \ref{deconfinedResultsSummary}.
\begin{table}[h!]
\centering
\begin{tabular}{|c||c|c|c|c|}
    \hline
      & $\Omega_d$ & $\Omega_e$ & $\Omega_f$ & $\Omega_g$\\
    \hline\hline
    $\Omega_d$ & $=0$ & $>0$ & $>0$ & $>~0$\\
    \hline
    $\Omega_e$ & $<0$ & $=0$ & - & -\\
    \hline
    $\Omega_f$ & $<0$ & - & $=0$ & phase transition\\
    \hline
    $\Omega_g$ & $<0$ & - & phase transition & $=0$\\
    \hline
\end{tabular}
\caption{Summary of deconfined phase results\ignore{ shown in Figures \ref{O20Deconfined}, \ref{O30Deconfined}, \ref{O05Deconfined}, \ref{O32Deconfd1is1}, \ref{O25Deconfd1is1}, \ref{O25Deconfd1ispt6}, \ref{O35Deconfd1is1} and \ref{O35Deconfd1ispt3}}. Results in various boxes show quantity in first row subtracted out of the quantity in first coloumn.}
\label{deconfinedResultsSummary}
\end{table}

\section{Discussion}\label{Discussion}

In this work, we have considered phases of $SU(N_c)$ gauge theory with large number of colours and two flavours at finite baryon and isospin chemical potential raised to finite temperature using Witten-Sakai-Sugimoto model. Each phase corresponds to a particular brane configuration obtained after introducing sources for $U(1)_B$ and $U(1)_3$ gauge fields in the bulk, which are 4-branes wrapped on $S^4$ part of the background spacetime and dissolved in flavour 8-branes. Temperature is introduced by using imaginary time formalism in the confined phase and through black brane background in the deconfined phase. Further, phase transitions are studied by comparing the actions of various configurations.

Let us first justify the assumption that the working temperature in the deconfined phase is larger than the deconfinement transition temperature. There are two spacetime metrics considered in this work. At low temperatures, the spacetime of equation (\ref{ConfB/g}) is energetically preferred over that of equation (\ref{DeconfB/g}), and the temperature is given by the inverse of the periodicity of $x_0$ circle. As temperature is increased, $x_0$ circle shrinks. At certain temperature, the period of $x_0$ circle becomes equal to that of the $x_4$ circle (which is written in equation (\ref{x4Period})), beyond which the spacetime of equation (\ref{DeconfB/g}) is favoured over that of equation (\ref{ConfB/g}). This phase transition \cite{HPAnalog4WSS} is analogous to Hawking-Page transition \cite{HawPaAdSBHPhTr} in asymptotic AdS spacetimes and corresponds to deconfinement transition in the gauge theory. Deconfinement transition temperature is given by $\frac{\beta_d}{R}=\delta x_4$, or $t_d = \frac{3}{4\pi}u_{KK}^{1/2}$. To ensure that the working temperature in the deconfined phase is larger than deconfinement temperature, one can go back to the confined phase and simply change $u_{KK}$ to make the $x_4$ circle larger, so that the circumference of the $x_0$ circle crosses that of the $x_4$ circle at a smaller temperature.

We have assumed that the deconfinement transition happens at a temperature lower than the temperature of chiral symmetry restoration. Reference \cite{0604161} shows that this is the case below a critical value of the asymptotic brane-anti-brane separation, $l$, implying that $l=1<l_{cr}$. To justify this, note that deconfinement temperature depends on $u_{KK}$ while chiral symmetry restoration temperature depends on $l$ (or $u_0$). In the case of $l=1$ being greater than $l_{cr}$, one can go back to the confined phase and change $u_{KK}$ such that the decconfinement transition happens at a lower temperature.


With isospin chemical potential in the theory, there are two ways for the D8-branes to separate. The configurations `b' and `e', shown in Figures \ref{confinedConfig2} and \ref{deconfinedConfig23}, have 8-branes crossing in the bulk. It has been shown \cite{CrossingBranes1,CrossingBranes2,CrossingBranes3} that such non-BPS intersecting brane configurations have a tachyonic instability. A series of papers \cite{CrossingBranesT1,CrossingBranesT2,CrossingBranesT3} discusses such tachyons in flat background and in Yang-Mills approximation. It was shown, using a finite temperature field theory calculation, that the tachyons disappear beyond a certain critical temperature ($T_{c}$), thereby stabilising the configuration. Accordingly, the configurations `b' and `e' do not exist for $T<T_c$.

In configuration `c' of Figure \ref{confinedConfig3}, the two sources are separated. The exact mechanism of this separation is not clear. However, this does represent a boundary theory with baryon number and isospin chemical potential. Ideally, one would like to consider energy required for separation of the sources while calculating the free energy difference. Here, we assume that the sources are separated by hand and leave further analysis for future.

On the gauge theory side, vacuum (which is the phase with no baryon number and isospin chemical potential, $d_0=d_3=0$, and corresponds to the U-shaped configurations of Figures \ref{confinedConfig0} and \ref{deconfinedConfig015}) is the least favoured, in both confined and deconfined phases, with comparison of all other phases with vacuum showing no phase transition. This is contrary to \cite{BLL}, which has phase transition between U-shaped and parallel brane configurations. However, in the present work, $\Omega_{dg}$ is explored by fixing $d_{1g}$ (see Figure \ref{O05Deconfined}). It is expected that changing the value of $d_{1g}$ further below $0.4$ will bring a phase transition between the phases D and G, though these values have not been explored.

Confined phase has chiral symmetry broken and there are two nuclear matter phases, B and C, corresponding to the two connected brane configurations, `b' (Figure \ref{confinedConfig2}) and `c' (Figure \ref{confinedConfig3}). Phase B has $U(2)$ symmetry broken \cite{BaSw} while phase C appears to have $U(2)$ symmetry restored in the UV. Configurations `b' and `c' have different UV asymptotics, accordingly phases B and C have not been compared. The deconfined phase also has two nuclear matter phases, E and F, with broken chiral symmetry. There is also a phase G, corresponding to the parallel brane configuration `g' (Figure \ref{deconfinedConfig015}), which has chiral symmetry restored. Comparing phases F and G shows phase transitions (Figures \ref{O35Deconfd1is1}, \ref{O35Deconfd1ispt3}, \ref{phaseFG}) with phase F favoured at lower temperatures and phase G favoured at higher temperatures, implying restoration of chiral symmetry at temperatures beyond this transition, as expected in QCD. This phase transition is on the same lines as the one between nuclear matter and QGP phases in \cite{BLL} and is expected because the corresponding brane configurations `f' and `g' are generalisations, to two pairs of D8-$\overline{\mbox{D8}}$, of the cusp and parallel-brane configurations respectively in \cite{BLL}. Phase E has not been compared with phases F and G on account of the corresponding brane configuration `e' having UV asymptotic behaviour different from that of `f' and `g'.

An important quantity is the grand potential function for a brane configuration. Though this is an infinite quantity, the corresponding difference for the two configurations is finite. To compare phases, we have plotted this difference, $\Delta\Omega$, vs $\mu_2$ (chemical potential for the second 8-brane for the two configurations), for various temperatures $t$ (in deconfined phase) and various values of $\mu_1$ (chemical potential for the first 8-brane for the two configurations). Ideally, one would like to have a three-dimensional plot, $t-\mu_1-\mu_2$. However, because of the complexity of expressions, we have considered only a few values of $\mu_1$ and $t$. Further, note that $\mu_1$ and $\mu_2$ are the chemical potentials for the two flavours in the boundary theory. To get back baryon number and isospin chemical potential ($\mu_0$ and $\mu_3$), the $\mu_1-\mu_2$ plane (in the $t-\mu_1-\mu_2$ plot) needs to be rotated about $t$ axis by $\frac{\pi}{4}$ radians.

To explore the phase transitions further, one can consider free energy as a function of an order parameter. This corresponds to writing the effective action as a function of some bulk field. Another possible direction is to calculate transport coefficients in this model.

\vspace{1cm}
\noindent
{\large {\bf Acknowledgments}}\\
The author would like to thank Swarnendu Sarkar for discussions and suggestions while the work was in progress, and Shigeki Sugimoto for some discussions prior to this work, which helped immensely in understanding the background. Gratitude is also expressed to Matthew Lippert, Swarnendu Sarkar and B. Sathiapalan for useful comments on the manuscript. Acknowledgement is due to Johanna Erdmenger, Giuseppe Policastro and Anton Rebhan for helpful conversations and to International Centre for Theoretical Sciences (ICTS), Bengaluru, India, for hospitality during the event `The Myriad Colorful Ways of Understanding Extreme QCD Matter' (code: ICTS/extremeqandg/2019/04) while this work was in progress. This work was supported by the Council of Scientific and Industrial Research (CSIR, under Ministry of Human Resource Development, Government of India), through grant 09/045(1355)/2014-EMR-I.

\newpage

\appendix

\section{Confined phase}

\subsection{Action}\label{action}

Action for embedding of $N_f$ D8-branes in the background generated by equataion (\ref{ConfB/g}) can be written using the non-Abelian generalisation of the DBI action \cite{TseyNADBI},
\begin{equation}
S^{D8} = \frac{1}{(2\pi)^8\alpha'^{9/2}} \int d^9X~e^{-\Phi}STr\sqrt{-det(g_{MN}+2\pi\alpha'{\cal F_{MN}})},
\end{equation}
with ${\cal F}$ being the $U(N_f)$ field strength and $g_{MN}$ the pull-back of the background spacetime. The $U(N_f)$ field can be decomposed as
\begin{equation}\label{NAgfdecomp}
 {\cal A} = \hat{A} + \frac{A}{\sqrt{2N_f}},
\end{equation}
with $\hat{A}$ being the $SU(N_f)$ field and $A$ the $U(1)$ field. As explained in section \ref{intro}, since the motivation is to study a theory with chemical potential in the boundary, we concern ourselves with the Abelian gauge field and use that in the DBI action.

\begin{equation}\label{S_D8Confined}
S^{D8} = C^{D8} \int du \left\{u^8 f(u) (x_4^\prime(u))^2 
 + u^5
    \left( {\frac{1}{f(u)}} - (a'_0(u))^2\right)
\right\}^{\frac{1}{2}},
\end{equation}
where $a_0'$ is the dimensionless Abelian gauge field from equation (\ref{Agfnorm}). To implement finite baryon density in the boundary theory, it is useful to Legendre transform away the $U(1)$ gauge field and work with the canonical conjugate,
\begin{equation}
\tag{\ref{d_confined}}
d(u) \equiv  \mbox{} - \frac{1}{C^{D8}} \frac{\partial {\cal L}^{D8}}{\delta a_0^\prime(u)}.
\end{equation}
This gives the Legendre transformed action in equation (\ref{LTS_D8Confined}).

Next, we evaluate the action for the 4-branes. Integrating equation (\ref{eomd}), one obtains
\begin{equation}\label{4branesDensity}
    n_4 = \frac{2\pi\alpha' R^2 C^{D8}} {\beta V_3 N_c} d.
\end{equation}
DBI action for $N_4$ of D4-branes described in section \ref{confined} is
$$
    S^{D4} = N_4\cdot\frac{1}{(2\pi)^4\alpha'^{5/2}} \int \left.d\Omega_4 d\tau e^{-\phi} \sqrt{g}\right|_{u=u_c},
$$
where $g$ is the determinant of the metric induced on these 4-branes courtesy background spacetime in equation (\ref{ConfB/g}), and is diagonal. Further, $N_4$ is related to $n_4$ as $n_4 = \frac{N_4}{V_3/R^3}$. The integral above evaluates to ${\frac{\Omega_4\beta R^4}{g_s}u_c}$, which gives (using equations (\ref{ConfB/g}) and (\ref{4branesDensity})),
\begin{equation}
   S^{D4} = \frac{1}{3}C^{D8}u_cd. \tag{\ref{D4onS4}}
\end{equation}

\subsection{D8-brane tension}\label{force}

Let the distance along the brane be denoted by $x$. Now,
$$
-\frac{1}{\sqrt{g_{uu}}}\frac{\partial \tilde{S}^{D8}}{\partial u_c} = C^{D8} u_c^4 \times \frac{1}{u_c^{3/2}\sqrt{f(u_c)}} \underbrace{\left\{1 + f(u_c)^2 u_c^3 (x'_4(u_c))^2\right\}^{1/2}}_{\frac{1}{cos \theta}} \times \left(1+\frac{d^2}{u_c^5} \right) \underbrace{u_c^{3/4}\sqrt{f(u_c)}}_{\frac{1}{\sqrt{g_{uu}}}}
$$
$$
\hspace{-5cm}= C^{D8} u_c^{13/4} \left(1+\frac{d^2}{u_c^5}\right)^{1/2} \frac{1}{cos \theta}
$$
$$
\hspace{-7cm}= \frac{1}{\sqrt{g_{uu}}}\frac{\partial \tilde{S}^{D8}}{\partial x \cdot cos \theta}.
$$
The proper tension along D8-brane is then given by
\begin{equation}
\frac{1}{\sqrt{g_{uu}}}\frac{\partial \tilde{S}^{D8}}{\partial x} = f^{D8} = u_c^{13/4} \left(1+\frac{d^2}{u_c^5}\right)^{1/2}.
\end{equation}

\subsection{Chemical potential}\label{chemicalPotential}
Shown next is the calculation for the chemical potential for the first D8-brane in configuration `c'.
\begin{eqnarray}
    \mu_{1_c} = \frac{1}{C^{D8}} \left\{
\int_{u_{c_1}}^\infty du\ \left(
\frac{\delta\tilde{\cal L}^{D8}_1}{\delta d_1(u)}  +
\frac{\delta\tilde{\cal L}^{D8}_1}{\delta x_{4_1}'(u)}
\frac{\partial x'_{4_1}}{\partial d_1} \right)\bigg|^{solution}_{t,l_1,l_2,u_{c_2}}\right.\nonumber\\
~~~~ \left.+ \frac{\partial u_{c_1}}{\partial d_1}\bigg|_{t,l_1,l_2}
 \left(\frac{\partial\tilde{S}_1^{D8}}{\partial u_{c_1}} +
\frac{\partial S^{D4}_1}{\partial u_{c_1}}\right)\right|^{solution}_{d_1,d_2,t,l_1,l_2}
+ \left.\frac{\partial S^{D4}_1}{\partial d_1}\bigg|_{t,l_1,l_2,u_{c_2}}
\right\}.
\end{eqnarray}
$\frac{\delta\tilde{\cal L}^{D8}_1}{\delta x_{4_1}'(u)}$ is constant, since there is no $x_{4_1}$ dependence in action equation (\ref{LTS_D8Confined}). Further, $\int_{u_{c_1}}^{\infty}du\frac{\delta x'_{4_1}}{\delta d_1}$ gives $-\frac{\delta l_1}{\delta d_1}$, which evaluates to zero at constant $l_1$. The sum $\left(\frac{\partial\tilde{S}_1^{D8}}{\partial u_{c_1}} +
\frac{\partial S^{D4}_1}{\partial u_{c_1}}\right)\bigg|^{solution}_{d_1,d_2,t,l_1,l_2}$ needs to vanish for the first D8-brane and the corresponding baryon vertex to be in equilibrium. This gives
\begin{equation}\label{chemPot}
    \mu_{1_c} = \int_{u_{c_1}}^\infty du~ a'_{1_0}(u) +  \frac{1}{C^{D8}} \left.\frac{\partial S_1^{D4}}{\partial d_1}\right|_{t,l_1,l_2,u_{c_2}}.
\end{equation}

$\mu_{2_c}$ can be calculated similarly. The corresponding quantities for configuration `b', $\mu_{1_b}$ and $\mu_{2_b}$, have $u_{c_1}=u_{c_2}$. Results are written in equations (\ref{chemPotConfinedConf2}) and (\ref{chemPotConfinedConf3}).

\newpage


\bibliographystyle{unsrt}
\bibliography{main}

\begin{thebibliography}{10}

\bibitem{HawPaAdSBHPhTr}
S.~W. Hawking and D.~N. Page.
\newblock {Thermodynamics of black holes in anti-de Sitter space}.
\newblock {\em Commun. Math. Phys.87(1983), 577}.

\bibitem{WiHolQCD1}
Edward Witten.
\newblock {Anti-de Sitter space, thermal phase transition and confinement in
  gauge theories}.
\newblock {\em Adv. Theor. Math. Phys. 2, 505 (1998) arXiv:hep-th/9803131}.

\bibitem{HPAnalog4WSS}
S.~Kalyana Rama and B.~Sathiapalan.
\newblock {The Hagedorn Transition, Deconfinement and the AdS/CFT
  Correspondence}.
\newblock {\em Mod.Phys.Lett. A13 (1998) 3137-3144 arXiv:hep-th/9810069}.

\bibitem{TseyNADBI}
A.~A. Tseytlin.
\newblock {On non-abelian generalisation of Born-Infeld action in string
  theory}.
\newblock {\em Nucl.Phys. B501 (1997) 41-52 arXiv:hep-th/9701125}.

\bibitem{KarchnKatz}
Andreas Karch and Emanuel Katz.
\newblock { Adding flavor to AdS / CFT}.
\newblock {\em JHEP 0206 (2002) 043 arXiv:hep-th/0205236}.

\bibitem{D4/D6}
M.~Kruczenski, D.~Mateos, R.~C. Myers, and D.~J. Winters.
\newblock {Towards a holographic dual of large-$N_c$ QCD}.
\newblock {\em JHEP 0405 (2004) 041 arXiv:hep-th/0311270}.

\bibitem{SaSu1}
T.~Sakai and S.~Sugimoto.
\newblock {Low energy hadron physics in holographic QCD}.
\newblock {\em Prog.Theor.Phys. 113, 843 (2005) arXiv:hep-th/0412141}.

\bibitem{SaSu2}
T.~Sakai and S.~Sugimoto.
\newblock {More on a Holographic Dual of QCD}.
\newblock {\em Prog.Theor.Phys. 114 (2005) 1083-1118 arXiv:hep-th/0412141}.

\bibitem{D3/D7}
J.~Babington, J.~Erdmenger, N.~Evans, Z.~Guralnik, and I.~Kirsch.
\newblock {Chiral symmetry breaking and pions in nonsupersymmetric gauge /
  gravity duals}.
\newblock {\em Phys.Rev. D69 (2004) 066007 arXiv:hep-th/0306018}.

\bibitem{D3/D7isospinCP}
Johanna Erdmenger, Matthias Kaminski, and Felix Rust.
\newblock {Isospin diffusion in thermal AdS/CFT with flavor}.
\newblock {\em Phys.Rev. D76 (2007) 046001 arXiv:0704.1290 [hep-th]}.

\bibitem{0B/YM4}
Roberto Grena, Simone Lelli, Michele Maggiore, and Anna Rissone.
\newblock {Confinement, asymptotic freedom and renormalons in type 0 string
  duals}.
\newblock {\em JHEP 0007 (2000) 005 arXiv:hep-th/0005213}.

\bibitem{D4D0/D8}
Shigenori Seki and Sang-Jin Sin.
\newblock {A New Model of Holographic QCD and Chiral Condensate in Dense
  Matter}.
\newblock {\em JHEP 1310 (2013) 223 arXiv:1304.7097 [hep-th]}.

\bibitem{hardwall}
Joshua Erlich, Emanuel Katz, Dam~T. Son, and Mikhail~A. Stephanov.
\newblock {QCD and a holographic model of hadrons}.
\newblock {\em Phys.Rev.Lett. 95 (2005) 261602 arXiv:hep-ph/0501128}.

\bibitem{softwall}
Andreas Karch, Emanuel Katz, Dam~T. Son, and Mikhail~A. Stephanov.
\newblock {Linear confinement and AdS/QCD}.
\newblock {\em Phys.Rev. D74 (2006) 015005 arXiv:hep-ph/0602229}.

\bibitem{ihqcd}
Umut Gürsoy, Elias Kiritsis, Liuba Mazzanti, Georgios Michalogiorgakis, and
  Francesco Nitti.
\newblock {Improved Holographic QCD}.
\newblock {\em Lect.Notes Phys. 828 (2011) 79-146 arXiv:1006.5461 [hep-th]}.

\bibitem{hish}
Jacob Sonnenschein.
\newblock {Holography Inspired Stringy Hadrons}.
\newblock {\em Prog.Part.Nucl.Phys. 92 (2017) 1-49}.

\bibitem{DLnHQCD}
Koji Hashimoto, Sotaro Sugishita, Akinori Tanaka, and Akio Tomiya.
\newblock {Deep Learning and Holographic QCD}.
\newblock {\em Phys.Rev. D98 (2018) no.10, 106014 arXiv:1809.10536 [hep-th]}.

\bibitem{0604161}
O.~Aharony, J.~Sonnenschein, and S.~Yankielowicz.
\newblock {A Holographic Model of Deconfinement and Chiral Symmetry
  Restoration}.
\newblock {\em Annals Phys. 322 (2007) 1420-1443 arXiv:hep-th/0604161}.

\bibitem{0608198}
Norio Horigome and Yoshiaki Tanii.
\newblock {Holographic chiral phase transition with chemical potential}.
\newblock {\em JHEP 0701 (2007) 072 arXiv:hep-th/0608198}.

\bibitem{similar}
Andrei Parnachev.
\newblock {Holographic QCD with Isospin Chemical Potential}.
\newblock {\em JHEP 0802 (2008) 062 arXiv:0708.3170 [hep-th]}.

\bibitem{WiBaryonVertex}
E.~Witten.
\newblock {Baryons And Branes In Anti de Sitter Space}.
\newblock {\em JHEP 9807 (1998) 006 arXiv:hep-th/9805112}.

\bibitem{BLL}
O.~Bergman, G.~Lifschytz, and M.~Lippert.
\newblock {Holographic Nuclear Physics}.
\newblock {\em JHEP {\bf 11}, 056 (2007) arXiv:0708.0326 [hep-th]}.

\bibitem{Rebhanreview}
Anton Rebhan.
\newblock {The Witten-Sakai-Sugimoto model: A brief review and some recent
  results}.
\newblock {\em EPJ Web Conf. 95 (2015) 02005 arXiv:1410.8858 [hep-th]}.

\bibitem{GubserTASI10}
Steven~S. Gubser.
\newblock {TASI lectures: Collisions in anti-de Sitter space, conformal
  symmetry, and holographic superconductors}.
\newblock {\em arXiv:1012.5312 [hep-th]}.

\bibitem{0908.0011}
Steven~S. Gubser, Silviu~S. Pufu, and Fabio~D. Rocha.
\newblock {Quantum critical superconductors in string theory and M-theory}.
\newblock {\em Phys.Lett. B683 (2010) 201-204 arXiv:0908.0011 [hep-th]}.

\bibitem{bottomupscd}
Sean~A. Hartnoll, Christopher~P. Herzog, and Gary~T. Horowitz.
\newblock {Building a Holographic Superconductor}.
\newblock {\em Phys.Rev.Lett. 101 (2008) 031601 arXiv:0803.3295 [hep-th]}.

\bibitem{BaSw}
S.~Kalyana Rama, S.~Sarkar, B.~Sathiapalan, and N.~Sircar.
\newblock {Strong Coupling BCS Superconductivity and Holography}.
\newblock {\em Nucl.\ Phys.\ B 852 (2011) 634 arXiv:1104.2843 [hep-th]}.

\bibitem{CrossingBranes1}
Satoshi Nagaoka.
\newblock {Higher Dimensional Recombination of Intersecting D-branes}.
\newblock {\em JHEP 0402 (2004) 063 arXiv:hep-th/0312010}.

\bibitem{CrossingBranes2}
F.~T.~J. Epple and D.~Lust.
\newblock {Tachyon condensation for intersecting branes at small and large
  angles}.
\newblock {\em Fortsch. Phys. 52 (2004) 367 [hep-th/0311182]}.

\bibitem{CrossingBranes3}
Niko Jokela and Matthew Lippert.
\newblock {Inhomogeneous tachyon dynamics and the zipper}.
\newblock {\em JHEP 0908 (2009) 024 arXiv:0906.0317 [hep-th]}.

\bibitem{CrossingBranesT1}
Sudipto~Paul Chowdhury, Swarnendu Sarkar, and B.~Sathiapalan.
\newblock {BCS Instability and Finite Temperature Corrections to Tachyon Mass
  in Intersecting D1-Branes}.
\newblock {\em JHEP 1409 (2014) 063 arXiv:1403.0389 [hep-th]}.

\bibitem{CrossingBranesT2}
Varun Sethi, Sudipto~Paul Chowdhury, and Swarnendu Sarkar.
\newblock {Finite Temperature Corrections to Tachyon Mass in Intersecting
  D-Branes}.
\newblock {\em JHEP 1704 (2017) 109 arXiv:1610.07140 [hep-th]}.

\bibitem{CrossingBranesT3}
Swarnendu Sarkar and Varun Sethi.
\newblock {Intersecting D-brane Stacks and Tachyons at Finite Temperature}.
\newblock {\em Nucl.Phys. B936 (2018) 118-150 arXiv:1801.02059 [hep-th]}.

\bibitem{Casalderrey-Solana1}
Maximilian Attems, Yago Bea, Jorge Casalderrey-Solana, David Mateos, Miquel
  Triana, and Miguel Zilhao.
\newblock {Holographic Collisions across a Phase Transition}.
\newblock {\em Phys.Rev.Lett. 121 (2018) no.26, 261601 arXiv:1807.05175
  [hep-th]}.

\bibitem{Casalderrey-Solana2}
Maximilian Attems, Yago Bea, Jorge Casalderrey-Solana, David Mateos, and Miguel
  Zilhao.
\newblock {Dynamics of Phase Separation from Holography}.
\newblock {\em arXiv:1905.12544 [hep-th]}.

\bibitem{JHEPReferee11}
Si~wen Li, Andreas Schmitt, and Qun Wang.
\newblock {From holography towards real-world nuclear matter}.
\newblock {\em Phys.Rev. D92 (2015) no.2, 026006 arXiv:1505.04886 [hep-ph]}.

\bibitem{JHEPReferee12}
Francesco Bigazzi and Aldo~L. Cotrone.
\newblock {Holographic QCD with Dynamical Flavors}.
\newblock {\em JHEP 1501 (2015) 104 arXiv:1410.2443 [hep-th]}.

\bibitem{JHEPReferee13}
Si~wen Li and Tuo Jia.
\newblock {Dynamically flavored description of holographic QCD in the presence
  of a magnetic field}.
\newblock {\em Phys.Rev. D96 (2017) no.6, 066032 arXiv:1604.07197 [hep-th]}.

\bibitem{JHEPReferee14}
Florian Preis, Anton Rebhan, and Andreas Schmitt.
\newblock {Inverse magnetic catalysis in dense holographic matter}.
\newblock {\em JHEP 1103 (2011) 033 arXiv:1012.4785 [hep-th]}.

\bibitem{JHEPReferee15}
E.~Antonyan, J.A. Harvey, S.~Jensen, and D.~Kutasov.
\newblock {NJL and QCD from string theory}.
\newblock {\em arXiv:hep-th/0604017}.

\end{thebibliography}

\end{document}